# Mid IR hollow core fiber gas laser emitting at 4.6 μm

F. B. A. Aghbolagh,[1,*] V. Nampoothiri,[1] B. Debord,[2] F. Gerome,[2]
L. Vincetti,[3] F. Benabid,[2] W. Rudolph[1]

[1]Department of Physics and Astronomy, University of New Mexico, Albuquerque, NM 87131, USA
[2] GPPMM group, Xlim Research Institute, UMR CNRS 7252, Université de Limoges, France
[3] Dept. of Engineering "Enzo Ferrari", University of Modena and Reggio Emilia, Italy
*Corresponding author: F. B. A. Aghbolagh_farzinbeygiazar2020@unm.edu

## 1. ABSTRACT

**Emission at 4.6 μm was observed from an $N_2O$ filled hollow core fiber laser. 8-ns pump pulses at 1.517 μm excited a vibrational overtone resulting in lasing on an R and P branch fundamental transition from the upper pump state. At optimum gas pressure of 80 Torr a photon conversion efficiency of 9% and a slope efficiency of 3% was observed from a mirrorless laser. The laser threshold occurred at an absorbed pump energy of 150 nJ in a 45-cm long fiber with 85 μm core diameter. The observed dependence of the laser output on gas pressure is shown to be a result of line broadening and relaxation rates.**

## 2. INTRODUCTION

Infrared laser sources emitting in the 3-5 μm and 8-13 μm atmospheric transmission window are attractive for applications such as remote sensing, imaging and free-space communications [1]. Extension of the emission wavelength of silica-based solid core fiber lasers into the IR spectral region is problematic because of absorption losses of the host materials. The advent of hollow core photonic crystal fibers (HCPCFs) [2] and the possibility to fill them with gases opened new avenues for laser development and nonlinear optics [3], taking advantage of the long interaction length of the interacting fields and their small modal areas. Within this context, the HCPCFs that guide via Inhibited Coupling (IC) [4] are attractive because they exhibit a guided field that overlaps weakly with the glass cladding material. The optical overlap factors $\eta$, which can range from $10^{-4}$ to $10^{-6}$ [5], are lower than those of index guiding solid core fibers or photonic bandgap HCPCFs. Consequently, material absorption losses are modest up to the mid infrared region for a few meter long IC guiding HCPCFs made from silica. IC guiding HCPCFs have already been demonstrated to guide beyond the 4 μm region where bulk fused silica becomes opaque [6,7].

For lasers, the useable transmission region can be expected to extend into the IR as long as the gain exceeds the propagation loss (per unit length), which is possible with gas active media. Gas filled HCPCFs have the guiding properties of photonic crystal structures while pushing the onset of detrimental nonlinear optical processes to higher intensities compared with solid core fibers [8,9]. Gas filled HCPCFs were reported first as both pulsed [10] and CW [11] Raman lasers, and later as pulsed systems based on population inversion in $C_2H_2$ [12]. The latter concept, hollow fiber gas laser (HOFGLAS), was also demonstrated as CW laser in the VIS with $I_2$ [13,14] and NIR with $C_2H_2$ [15]. A HCPCF filled with molecular hydrogen extended the output wavelength of a Raman laser to 4.4 μm [16].

Depending on the gas and spectral region of interest, the pump and gain transitions in HOFGLAS can involve electronic, vibrational and rotational states and combinations thereof. Transition cross sections range from >$10^{-14}$ cm$^2$ for fundamental, dipole allowed electronic transitions to < $10^{-21}$ cm$^2$ for excitation of vibrational overtones where the change of the vibrational quantum number $\Delta v \geq 2$. The optical density can be controlled by the gas pressure and fiber length. Compared to dopants in solid core fibers gases have much narrower line widths. However, the multitude of rotational levels provides the possibility of lasing at discrete wavelengths.

In this letter we report on an optically pumped $N_2O$ filled hollow core fiber laser pumped on a second overtone transition at 1.517 μm, which resulted in emission from a fundamental ro-vibrational transition ($3v_3$ to $2v_3$) at 4.6 μm. Lasing at 4.6 μm of $N_2O$ in a regular gas cell was achieved previously by using an OPO pump source at 2.88 μm [17]. Our system extends the output of HOFGLAS to the 4-5 μm region.

## 3. EXPERIMENTAL SETUP AND RESULTS

An energy level diagram of $N_2O$ with the relevant transitions is shown in Fig.1.

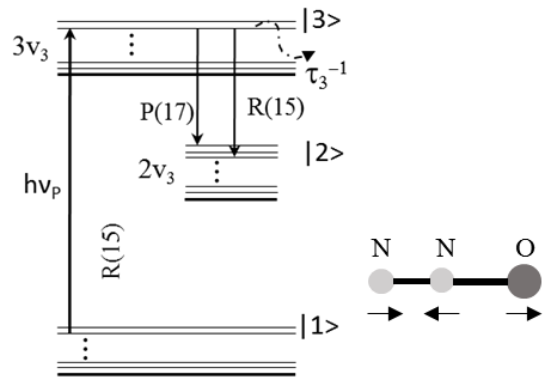

Fig.1. Simplified energy diagram of $N_2O$ showing three vibrational bands belonging to the N-O stretch mode $v_3$. The pump excites the second overtone.

Figure 2 compares the transmission behavior of bulk fused silica (solid data points) and an IC guiding HCPCF with 85 μm inner core diameter. The fiber distinguishes itself by its ultra-broad transmission windows and an extremely small optical overlap of its core mode $HE_{11}$ with the silica cladding. The latter property allows optical guidance in wavelength regions where the glass strongly absorbs

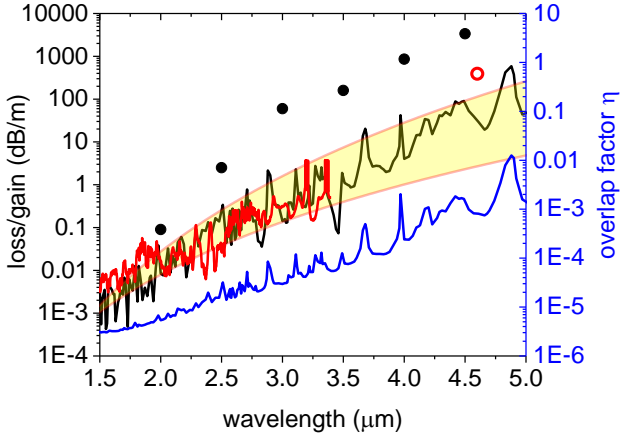

Fig.2. Loss coefficient $\kappa_F$ versus wavelength for bulk fused silica (solid data points, from Heraeus Suprasil 300 [18]). Measured loss in the HCPCF used for our HOFGLAS (red solid line). Numerically calculated confinement loss of the HCPCF core-fundamental mode (black solid line). The shaded region is a guide for the eye of the transmission loss range. The open red circle corresponds to the calculated gain coefficient of $N_2O$ at 4.6 μm, see text. The blue solid curve shows the calculated optical overlap of the fiber core fundamental mode.

The fiber exhibits a Kagome-lattice cladding and a hypocycloid (i.e. negative curvature) core-contour to enhance the IC and thus reducing the fiber confinement loss [19,20]. The fiber has an outer diameter of 300 μm, an inner core diameter of 85 μm, silica strut thickness of 490 nm, and a negative curvature parameter $b \approx 0.4$. The measured loss spectrum is shown from 1.5 μm to 3.4 μm (red solid line). At the pump wavelength (1.517 μm) the fiber loss is about 30 dB/km. The figure shows the calculated confinement loss (CL) of the core fundamental mode of the HCPCF over the wavelength range from 1.5 μm to 5 μm. The CL can also be estimated qualitatively from simple scaling laws reported in [21]. The calculated CL and the measured loss show a good agreement with the experimental data over the measured spectral range. For an overlap factor $\eta \ll 1$, the effective propagation loss coefficient can be estimated with the sum $\alpha_{eff} \approx \alpha_{CL} + \alpha_m \eta$, where $\alpha_m$ is the absorption coefficient of the cladding material (fused silica). Here, we ignore the loss due to surface scattering [5]. Figure 2 shows that the contribution of the material absorption to the propagation loss is ~3.6 dB/m, yielding an effective propagation loss of $\alpha_{eff} \approx$ 33.6 dB/m, which is more than one order of magnitude lower than the calculated $N_2O$ gain.

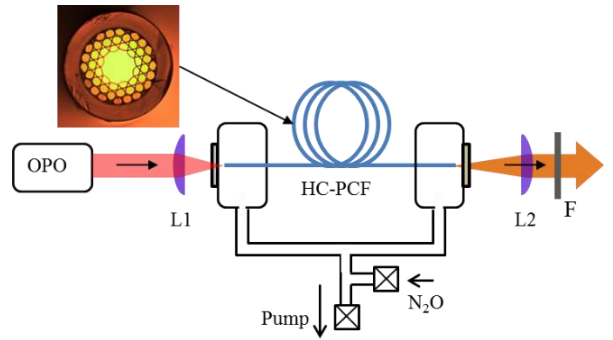

Fig.3. Experimental setup of the 4.6 μm pulsed $N_2O$ laser. The inset shows the cross section of the 45-cm long fiber. The pump was an 8-ns pulse at 1.517 μm from an OPO. To control the pressure the two ends of the fiber were housed in small chambers. "F" denotes a long pass germanium filter used to block residual pump radiation.

The experimental setup of the laser is shown in Fig.3. Pulses from an 8-ns OPO at 1.517 μm are coupled into a 45-cm long Kagome HCPCF using an uncoated $CaF_2$ lens with a focal length of 5-cm. The coupling efficiency was about 45%. The output is monitored with a spectrometer and a HgCdTe energy meter.

Lasing is observed without an external cavity. As reported with other HOFGLASs, the gain is large enough to produce laser-like output based on amplified spontaneous emission. The laser spectrum is shown in Fig.4. The two spectral lines of the laser emission belong to the P(17) and R(15) transition, originating at the upper pump level, see Fig.1.

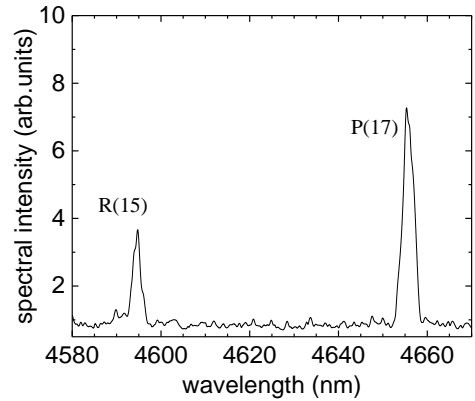

Fig.4. Measured spectrum of the $N_2O$ laser. Details of the lines are not resolved. $p$ = 80 Torr, absorbed pump energy $E_{abs}$ = 1.3 μJ.

Figure 5 shows the measured laser output as a function of the gas pressure. Maximum output was observed at a pressure of $p_{opt} \approx$ 80 Torr. Figure 6 shows the laser pulse energy as a function of the absorbed pump pulse energy. At optimum pressure the slope efficiency was 3%. The threshold occurred at an absorbed pump pulse energy of about 150 nJ.

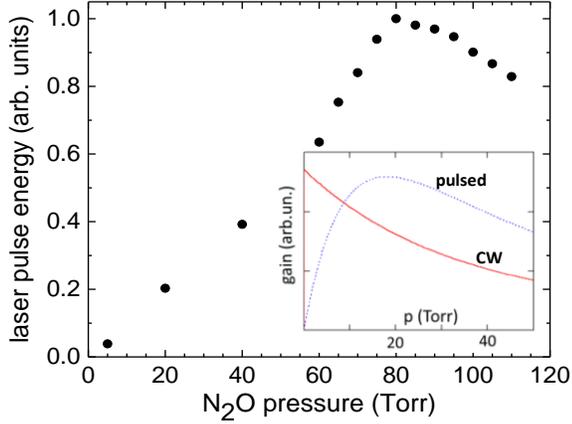

Fig.5. Measured laser output as a function of N$_2$O pressure for 70 µJ of pump energy coupled into fiber. The inset shows the calculated gain coefficient $g_{32}(p)$, see text.

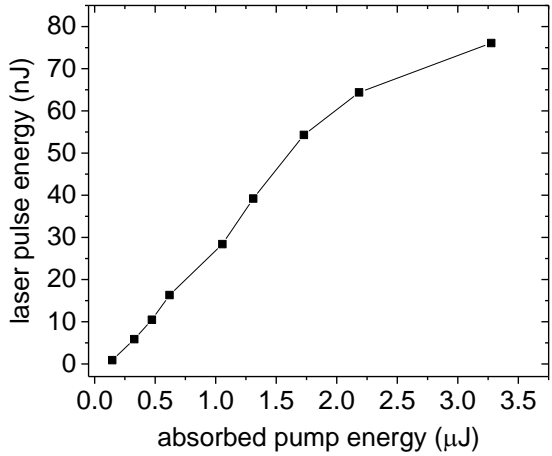

Fig.6. Laser pulse energy as a function of absorbed pump pulse energy for a 45-cm long fiber filled with 80 Torr of N$_2$O gas.

For an estimation of the gain and its pressure dependence we refer to the energy level scheme shown in Fig.1. For simplicity we will consider only one of the two possible gain transitions. To obtain lasing without feedback from a cavity the gain has to exceed the losses in each segment of the fiber if we assume constant pump intensity and neglect saturation. To obtain the gain coefficient we first calculate the number density of molecules excited to the upper laser level, N$_3$, if the pump (intensity $I_p$) is turned on at $t = 0$ and obtain:

$$N_3(t) = \frac{\sigma_{13} N_1}{h\nu_p} I_p \tau_3 (1 - e^{-t/\tau_3}) \quad (1)$$

Here $\sigma_{13}$ is the absorption cross section, $N_1$ is the number density of molecules in the rotational state (J = 15) of the vibrational ground state that absorbs the pump, and $1/\tau_3$ is the total removal rate of molecules from the excited rotational state, see Fig.1. The small signal gain coefficient

$$g_{32} = \sigma_{32} N_3, \quad (2)$$

with $\sigma_{32}$ being the gain cross section. We omitted a factor of approximately ½ that takes into account gain sharing between the R and P transition. For pulse durations $\tau_p \gg \tau_3$ and CW excitation, $g_{32} \approx \sigma_{32}\sigma_{13} N_1 I_p \tau_3 / h\nu_p$, and for $\tau_p \ll \tau_3$ the maximum gain coefficient reached at the end of the pulse $g_{32} \approx \sigma_{32}\sigma_{13} N_1 I_p \tau_p / h\nu_p$. These expressions were obtained after simplifying the term in brackets in Eq. (1). The second-overtone absorption cross section $\sigma_{13}$ for N$_2$O is about $10^{-20}$ cm$^2$. A typical emission cross section for a fundamental ro-vib transition in N$_2$O is $10^{-15}$ cm$^2$ [22]. At 300 K, about 4% of the N$_2$O molecules occupy the lower pump state. The relaxation time $\tau_3 = R/p$ with $R \approx 10^{-7}$ sTorr [23] as a typical value at room temperature for rotational relaxation. Thus, for a gas pressure of $p$ = 80 Torr and a 40 µJ pump pulse propagating in a HCPCF with 85 µm inner core diameter, $\tau_3 \approx 1.3$ ns and $N_3 \approx 9 \times 10^{14}$ cm$^{-3}$. Equation (2) then yields a gain coefficient of $g_{32} \approx 0.9$ cm$^{-1}$, which is shown in Fig. 2 (in dB/m).

As long as the fiber loss coefficient $\kappa_F < g_{32}$, spontaneously emitted light will be amplified. If the overall gain is large enough the radiation at the fiber end is spectrally narrowed with properties similar to a laser with external cavity. The relatively large gain coefficient suggests that HOFGLAS can extend even further into the IR, see Fig.2.

In HOFGLAS, the gas pressure is an important adjustment parameter for optimizing the laser output [9]. This results from the pressure dependence of the gas parameters that enter the expression for the gain; – transition line width $\Delta\nu$, cross sections $\sigma \propto 1/\Delta\nu$, relaxation rates $\tau_3 \propto R/p$ and number density in the ground state $N_1$. In the pressure and temperature range of interest, the transitions are affected by Doppler and collisional broadening and the width of the resulting Voigt profile can be approximated by $\Delta\nu(p) \approx 0.53\beta p + \sqrt{0.217\beta^2 p^2 + \Delta\nu_D^2}$ [24], where $\beta$ is the pressure broadening coefficient for self-collisions, and $\Delta\nu_D$ is the Doppler linewidth. For an estimation we use $\beta \approx 8$ MHz/Torr [22] for both absorption and gain transition and $\Delta\nu_D$ = 100 MHz, 370 MHz for the gain and absorption transition, respectively. The pressure dependence $g_{32}(p)$ can now be evaluated with the help of Eqs. (1) and (2) and is shown as inset in Fig.5 for a CW pump and an 8-ns pump pulse.

In both cases there is an optimum pressure that maximizes $g(p)$ for given input pump intensity. The optimum pressure value for CW is in the sub Torr region as observed in Ref. [25]. The reason is the fast removal rate from the excited state, $1/\tau_3$ for these conditions. For the pulsed case, the calculated optimum pressure $p_{opt}$ is lower than what we observed, 18 Torr versus 80 Torr. The discrepancy is due to a number of factors. First, the spectral width of our pump pulse $\Delta\nu_p \sim 3$ GHz, which leads to an increase in absorption with increasing pressure until about $p \approx 350$ Torr at which $\Delta\nu \sim \Delta\nu_p$. Second, the laser output is affected by saturation of the laser transition, which is neglected in the estimation for the small signal gain. A fast relaxation out of the lower laser level is beneficial for efficiency. This favors a larger pressure and is expected to increase $p_{opt}$.

## 4. SUMMARY

In summary, we have demonstrated an $N_2O$ HOFGLAS in the mid-infrared spectral region optically pumped on a second overtone. The achieved slope efficiency at optimum pressure of 80 Torr was 3%. No attempt was made to optimize the laser output by finding the optimal fiber length and other parameters. Optimization of these parameters is likely to result in higher slope efficiency. The observed emission wavelength of 4.6 μm and our gain estimations suggest the potential of silica based HOFGLASs to extend laser emission beyond 5 μm with gases as active medium.

**FUNDING.** Air Force Research Lab (FA9451-17-2-0011), College of Arts and Sciences at the University of New Mexico (UNM).

**ACKNOWLEDGMENT**. We are grateful for the loan of a diode laser under the Education Partnership Agreement (EPA) 2011-AFRL/RD&RV-EPA-01 between the Air Force Research Laboratory and the University of New Mexico. We thank L. Emmert (UNM), and K. Corwin and B. Washburn (Kansas State University) for helpful discussions.